\documentclass[prl,aps,showpacs,twocolumn]{revtex4}
\usepackage{bm,graphicx}
\usepackage{latexsym}
\begin{document}

\title{Quantum Phase Transition of Ground-state Entanglement in a Heisenberg Spin
Chain Simulated in an NMR Quantum Computer}
\author{Xinhua \surname{Peng}$^{1}$ } 
\author{Jiangfeng \surname{Du}$^{2,3}$ } 
\author{Dieter \surname{Suter}$^{1}$ }
\affiliation{$^{1}$Universit\"{a}t Dortmund, Fachbereich Physik,
44221 Dortmund, Germany}
\affiliation{$^{2}$Department of Physics, National University of Singapore, 117542
Singapore}
\affiliation{$^{3}$Hefei National Laboratory for Physical Sciences at Microscale and
Department of Modern Physics, University of Science and Technology of
China, Hefei, Anhui 230026, P.R. China}
\date{\today}

\begin{abstract}
Using an NMR quantum computer, we experimentally simulate the quantum phase
transition of a Heisenberg spin chain. The Hamiltonian is generated by a
multiple pulse sequence, the nuclear spin system is prepared in its
(pseudo-pure) ground state and the effective Hamiltonian varied in such a
way that the Heisenberg chain is taken from a product state to an entangled
state and finally to a different product state.
\end{abstract}


\maketitle

\draft

Quantum mechanical systems are known to undergo phase transitions at zero
temperature when a suitable control parameters in its Hamiltonian is varied 
\cite{Sacbook}. At the critical point where the quantum phase transition
(QPT) occurs, the ground state of the system undergoes a qualitative change
in some of its properties \cite{Sacbook}. Osterloh et al. \cite{Ost02}
showed that in a class of one-dimensional magnetic systems, the QPT is
associated with a change of entanglement, and that the entanglement shows
scaling behavior in the vicinity of the transition point. This behavior was
discussed in detail for the Heisenberg model \cite{Osb02} and for the
Hubbard model \cite{Gu04}. It is believed that the ground-state entanglement
also plays a crucial role in other QPTs, like the change of conductivity in
the Mott insulator-superfluid transition \cite{Gebbook} and the quantum Hall
effect \cite{Lau83}. Many of the relevant features, like the transition from
a simple product state to a strongly entangled state, occur over a wide
range of parameters and persist for infinite systems as well as for systems
with as few as two spins \cite{AZW,Kam02}. These systems, especially the
Heisenberg spin model, are central both to condensed-matter physics and to
quantum information theory. In quantum information processing, the
Heisenberg exchange interaction has been shown to provide a universal set of
gates \cite{Ima99,Rau01} and in quantum communication, information can be
propagated through a Heisenberg spin chain \cite{Bos03}.

While some Heisenberg models can be solved analytically, others can only be
simulated numerically. Like for other quantum systems, such simulations are
extremely inefficient if the system contains more than 10-20 spins. It was
therefore suggested that such simulations could be more efficiently
performed on a quantum computer \cite{Fey82}. In this Letter, we discuss the
simulation of a Heisenberg spin chain by an NMR quantum computer. By varying
the strength of the magnetic field, we take the system, which is in the
quantum mechanical ground state, through the QPT and measure the change in
entanglement by quantum state tomography. The NMR techniques that we use
here are closely related to earlier work where they were used to demonstrate
quantum algorithms, quantum error correction, quantum simulation, quantum
teleportation and more(see, e.g., Ref. \cite{NMRQC} and references cited
therein).

The simplest system that exhibits this behavior consists of two spins
coupled by the Ising interaction 
\begin{equation}
H = \frac{\omega_{z}}{2} (\sigma^1_z+\sigma^2_z) + J_I\sigma^1_z \sigma^2_z,
\quad\quad J_I > 0 ,
\end{equation}
where $\sigma^i_z$ are the Pauli operators, $\omega _{z}$ a magnetic field
strength, and $J_I$ is a spin-spin coupling constant.

In the range $-J_I\leq \omega _{z}\leq J_I$, the ground state of this system
is two-fold degenerate. To avoid this complication, we add a small
transverse magnetic field. The resulting Hamiltonian is 
\begin{equation}
H=\frac{\omega _{z}}{2}(\sigma _{z}^{1}+\sigma _{z}^{2})+\frac{\omega _{x}}{2%
}(\sigma _{x}^{1}+\sigma _{x}^{2}) + J_I\sigma _{z}^{1}\sigma _{z}^{2},
\label{e.Ham}
\end{equation}
which is nondegenerate. The transverse field will always be kept small, $%
|\omega _{x}|\ll |\omega _{z}|,|J_I|$.

A symmetry-adapted basis that is an eigenbasis for vanishing transverse
field ($\omega_x = 0$) is \{$\left| \uparrow \uparrow \right\rangle ,\left|
\Psi ^{+}\right\rangle ,\left| \downarrow \downarrow \right\rangle ,\left|
\Psi ^{-}\right\rangle \}$, with $\left| \Psi ^{\pm }\right\rangle =\frac{1}{%
\sqrt{2}}(\left| \uparrow \downarrow \right\rangle \pm \left| \downarrow
\uparrow \right\rangle )$ and $\left|\downarrow \right\rangle$ and $\left|
\uparrow \right\rangle$ the spin-down ($m=-\frac{1}{2}$) and spin up ($m=+%
\frac{1}{2}$) states. Furthermore, it is convenient to define dimensionless
field strengths $g_x = \frac{\omega_x}{2J}$ and $g_z = \frac{\omega_z}{2J}$.

Since the ground state of this system is always one of the triplet states,
and transitions to the singlet state are symmetry-forbidden, we can reduce
our system of interest to the triplet states. For small transverse fields, $%
g_x \ll 1$, the longitudinal field $g_z$ determines the ground state 
\begin{equation}
\left| \psi _1\right\rangle \simeq \left\{ 
\begin{array}{l}
\left| \uparrow \uparrow\right\rangle \hspace{0.7 cm} g_z < -1 \\ 
\left| \Psi^{+}\right\rangle \hspace{0.7 cm} -1 < g_z < 1 \\ 
\left| \downarrow\downarrow\right\rangle \hspace{0.7 cm} g_z > 1
\end{array}
\right. .
\end{equation}
$g_z = \pm 1$ are therefore quantum critical points, where the ground state
changes from the ferromagnetically ordered high field states to the
entangled, antiferromagnetic low-field state.

For the full system, including the transverse field, the eigenstates and
eigenvalues of the three-state system are

\begin{equation}
\begin{array}{l}
|\psi _{i}\rangle =\frac{1}{\sqrt{M_{i}}}(\frac{\xi _{i}^{2}+2(\xi
_{i}+1)g_{z}-1-2g_{x}^{2}}{2g_{x}^{2}}\left| \uparrow\uparrow\right\rangle  \\ 
+\frac{\xi _{i}-1+2g_{z}}{\sqrt{2}g_{x}}\left| \Psi ^{+}\right\rangle
+\left| \downarrow\downarrow\right\rangle ),(i=1,2,3)
\end{array}
\end{equation}
and 
\begin{equation}
\begin{array}{l}
{\large \varepsilon }_{1,2}=J_{I}\xi _{1,2}=J_{I}(1-2r\cos (\theta \mp \pi
/3))/3 \\ 
{\large \varepsilon }_{3}=J_{I}\xi _{3}=J_{I}(2r\cos \theta +1)/3,
\end{array}
\end{equation}
where $M_{i}$ are normalization constants, $r=2\sqrt{%
3(g_{x}^{2}+g_{z}^{2})+1}$, and $\theta =\frac{1}{3}\arccos (\frac{%
4(18g_{z}^{2}-9g_{x}^{2}-2)}{r^{3}})$.

\begin{figure}[tbh]
\begin{center}
\includegraphics[width = \columnwidth]{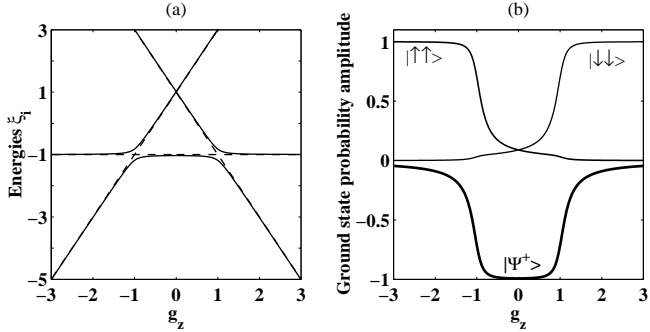}
\end{center}
\caption{(a) Energy level diagram for the two-spin Heisenberg Ising model
for $g_{x}=0$ (dashed lines) and $g_{x}=0.129$ (solid lines). (b)
Probability amplitudes of the ground state $\left| \psi _{1}\right\rangle$
for $g_{x}=0.129$.}
\label{fig:energy}
\end{figure}
Figure \ref{fig:energy} shows numerical values for the energies and the
coefficients of the ground state as a function of the longitudinal field
strength $g_{z}= \frac{\omega _{z}}{2J_I}$. The right hand side shows
clearly that at strong fields ($|g_z|>1$), the ground state is a product
state, while it corresponds to the entangled state $|\Psi ^+\rangle$ for
weak fields.

To observe the system undergoing the QPT, we simulate it on an NMR quantum
computer, where the quantum spins $\sigma^i$ are represented by nuclear
spins and the Hamiltonian (\ref{e.Ham}) of the Heisenberg chain is mapped
into an effective Hamiltonian generated by a sequence of radio frequency
(rf) pulses acting on the nuclear spin system.

The natural Hamiltonian of our two-qubit system is 
\begin{equation}
H_{NMR}=\frac{\omega^1_L}{2}\sigma _{z}^{1} + \frac{\omega^2_L}{2}\sigma_z^2
+ \frac{J_{12}}{4}\sigma _{z}^{1}\sigma _{z}^{2} .
\end{equation}
The $\omega^{1,2}_L$ represent the Larmor frequencies of the two qubits and $%
J_{12}$ the spin-spin coupling constant. In addition to this static
Hamiltonian, we use rf pulses to drive the dynamics of the system. In the
usual rotating coordinate system, the effect of rf pulses can be written as 
\begin{equation}
H_{rf} = \frac{\omega_{rf}}{2}(\sigma _{x}^{1} + \sigma_x^2)
\end{equation}
where we assumed that the rf field strength is the same for both qubits.

\begin{figure}[htb]
\begin{center}
\includegraphics[width = \columnwidth]{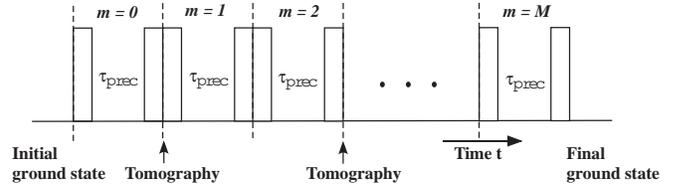} .
\end{center}
\caption{Sequence of rf pulses applied to both spins to simulate the target
Hamiltonian. The boxes represent pulses that induce rotations around the
x-axes of the rotating frame, the separations between them the free
precession periods. The index $m$ for the different periods runs from $0$ to 
$M$.}
\label{fig:f2}
\end{figure}

The target Hamiltonian (\ref{e.Ham}) can be created as an average
Hamiltonian by concatenating small flip angle rf pulses with short periods
of free evolution, $e^{-iH\tau }\simeq e^{-iH_{rf}\tau _{p}}e^{-iH_{NMR}\tau
_{prec}}$, where $\tau_p$ is the pulse duration and $\tau_{prec}$ the length
of the free evolution period. The resulting effective Hamiltonian matches
the target Hamiltonian if $\omega^1_{L}=\omega^2_{L}=\frac{\tau }{\tau
_{prec}}\omega _{z}$, $\tau _{p}=\frac{\omega _{x}}{\omega _{rf}}\tau $, and 
$\tau _{prec}=\frac{4 J_I} {J_{12}}\tau $. While this approximation is
correct to first order in $\tau $, the symmetrized sequence 
\begin{equation}
(H_{rf},\frac{\tau _{p}}{2})-(H_{NMR},\tau _{prec})-(H_{rf},\frac{\tau _{p}}{%
2})  \label{e.sequence}
\end{equation}
generates the desired evolution to second order in $\tau $. Figure \ref
{fig:f2} shows the sequence of rf pulses required to generate this evolution.

To prepare the system in the ground state, we use the technique of
pseudo-pure states \cite{pps}: we prepare a density operator $%
\rho_{pp}(\psi_1) = \frac{1}{tr(\mathbf{1})} \mathbf{1} + \alpha
|\psi_1\rangle\langle \psi_1|$. Here, $\mathbf{1}$ is the unity operator and 
$\alpha$ a small constant of the order of $10^{-5}$. To measure the order
parameter (entanglement), we apply quantum state tomography \cite{qst}. The
system can then be taken through the phase transition by adiabatically
changing the magnetic field $g_z$ of the effective Hamiltonian, which acts
as a control parameter.

To ensure that the system always stays in the ground state, the variation of
the control parameter has to be sufficiently slow, so that the condition $%
\left| \frac{\langle \psi _{1}(t)\left| \dot{\psi}_{e}(t)\right\rangle }{%
\varepsilon _{e}(t)-\varepsilon _{1}(t)}\right| \ll 1$ is fulfilled, where
the index $e$ refers to the excited states \cite{Messbook}. Choosing $g_{z}$
as the control parameter, we write the adiabaticity condition as 
\begin{equation}
\left| \frac{dg_{z}}{dt}\right| \ll J_{I}^{2}\chi =J_{I}^{2}\left| \frac{%
(\xi _{2}(t)-\xi _{1}(t))^{2}}{\langle \psi _{1}(t)|\frac{\partial H}{%
\partial g_{z}}|\psi _{2}(t)\rangle }\right| ,  \label{e.adcond}
\end{equation}
where the dimensionless parameter $\chi $ quantifies the sensitivity to the
control parameter $g_{z}$ and we have concentrated on the first excited
state $|\psi _{2}\rangle $, which is the critical one for transitions from
the ground state.

\begin{figure}[tbh]
\begin{center}
\includegraphics[scale = 0.29]{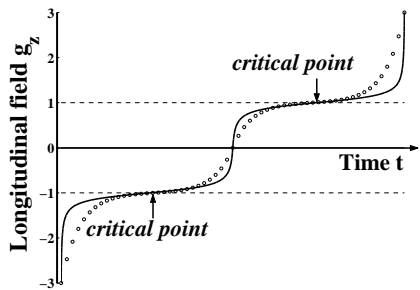}
\end{center}
\caption{Adiabatic magnetic field sweeps $g_{z}(t)$. The solid line was
calculated for constant adiabaticity parameter $\frac{\chi }{dg_{z}/dt}$
(see equ. (\ref{e.adcond})) for a transverse field $g_{x}=0.129$; the
circles represent the values obtained from the numerical optimisation of the
discretised scan.}
\label{f.sweep}
\end{figure}

Equation (\ref{e.adcond}) defines the optimal sweep of the control parameter 
$g_{z}(t)$, with the scan speed $\left| \frac{dg_{z}}{dt}\right| \propto
\chi $. Figure \ref{f.sweep} shows the required time dependence of the
magnetic field. The resulting transfer is therefore highest for a given scan
time or the scan time minimised for a required adiabaticity.

The experimental implementation generates an effective Hamiltonian that is
constant for a time $\tau$ (see equ. (\ref{e.sequence})). For this stepwise
approximation, the duration of each time step has to be chosen such that (i)
the time is short enough that the average Hamiltonian approximation holds
and (ii) the adiabaticity criterion remains valid. While this calls for many
short steps, there is also a lower limit for the duration of each step,
which is dictated by experimental aspects: switching transients, which are
not taken into account in the Hamiltonian of equ. (\ref{e.sequence}), tend
to generate errors that increase with the number of cycles.

We used a numerical optimisation procedure to determine the optimal sequence
of Hamiltonians, taking the full level structure into account. Choosing a
hyperbolic sine as the functional form, we optimised its parameters and
found the optimised discrete scan represented by the circles in Fig. \ref
{f.sweep}.

\begin{figure}[tbh]
\begin{center}
\includegraphics[width = 5 cm]{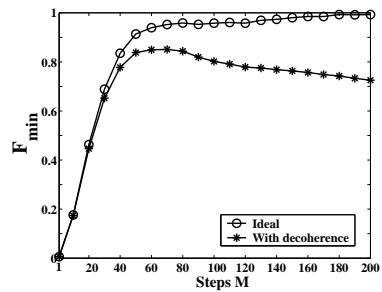} .
\end{center}
\caption{Numerical simulation of the minimum fidelities during the adiabatic
passage vs. number of steps with ($*$) and without ($\bigcirc $) decoherence
effects.}
\label{fig:f3}
\end{figure}

To determine the optimal number of steps, we used the same numerical
simulation, keeping the functional dependence $g_{z}$ vs. $t$ constant, but
increasing the number of steps. The results are summarized in Fig. \ref
{fig:f3}, which plots the lowest fidelity encountered during each scan
against the number of steps taken in the simulation. The fidelity is
calculated as the overlap of the state with the ground state at the relevant
position. The simulation shows also the effect of decoherence, which reduces
the achievable fidelity if the total duration of the scan becomes comparable
to the decoherence time. The model that we used to take the effect of
decoherence into account is similar to that of Vandersypen et al \cite{shor}.

For the experimental implementation, we used the $^{13}$C and $^{1}$H spins
of $^{13}$C\_labeled chloroform (both spins 1/2). The relatively large
spin-spin coupling constant of $\frac{J_{12}}{2\pi}$=214.94Hz makes this
molecule well suited for this experiment. The chloroform was diluted in
acetone-d$_6$ and experiments were carried out at room temperature on a
Bruker DRX-500 MHz spectrometer. The pseudo-pure initial state $%
\rho_{pp}(\uparrow \uparrow)$ was generated by spatial averaging \cite{pps}.
The fidelity of this state preparation was checked by quantum state
tomography and found to be better than 0.99.

The adiabatic scan through the QPT was achieved by shifting the rf
frequencies of both channels by the same amount after each period. Using the
sweep $g_z(t)$ shown in Fig. \ref{f.sweep}, the offset was changed from $%
g_z=-3$ to $g_z = +3$ in 60 steps. The evolution of the system during the
scan was checked by performing a complete quantum state tomography after
every second step during the experiment.

As a quantitative measure of the QPT, we used the concurrence as the order
parameter, which is related to "the entanglement of formation" \cite
{Wootters} and ranges form 0 (no entanglement) to 1 (maximum entanglement).
For this purpose, we calculated the concurrence \textit{C} from the
tomographically reconstructed deviation density matrices $\rho$ as $C(\rho
)=max\{\lambda _1-\lambda _2-\lambda_{3}-\lambda _{4},0\}$, where $%
\lambda_{i}(i=1,2,3,4)$ are the square roots of the eigenvalues of $\rho
(\sigma^1_{y} \sigma^2_{y})\rho ^{*}(\sigma^1_{y}\sigma^2_{y})$ in
decreasing order.

\begin{figure}[tbh]
\begin{center}
\includegraphics[scale = 0.45]{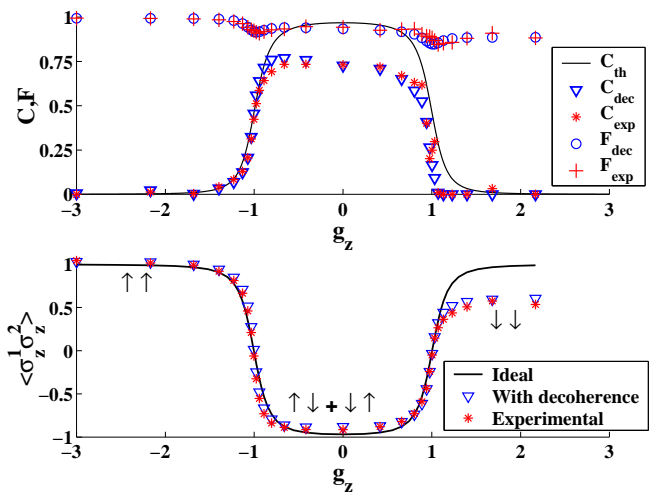} .
\end{center}
\caption{(Color online)(a) Measured fidelity $F_{\exp }$ ($+$) and concurrence $C_{\exp }$
($*$) compared to the concurrence calculated for an ideal scan $C_{th}$
(solid line) and the simulated concurrence $C_{dec}$ ($\bigtriangledown $)
and fidelity $F_{dec}$ ($\bigcirc $) when decoherence is taken into account.
(b) Measured values ($*$) of the two-spin correlaror $\left\langle \sigma
_{z}^{1}\sigma _{z}^{2}\right\rangle $ compared to the theoretical (solid
line) and simulated values with decoherence($\bigtriangledown $). }
\label{fig:f4}
\end{figure}

Figure \ref{fig:f4}(a) shows the measured concurrence $C_{\exp }=C(\rho
_{\exp})$ as individual points and compares them with the theoretical values 
$C_{th} $. Both data sets clearly show the expected QPTs near the critical
points $g_{z}=\pm 1$. The entangled ground state for $|g_z| < 1$ is
characterised by a concurrence close to 1, while the high field states are
only weakly entangled (the entanglement vanishes for $g_x = 0$).

The experimentally determined concurrence remains below $\sim 0.75$,
significantly less than the theoretical values. To verify that this
deviation is due to decoherence, we simulated the experiment, taking into
account the details of the pulse sequence as well as the effect of
decoherence. We obtained good agreement between theoretical and experimental
data if we assumed a total decoherence time of 130 ms, which is slightly
longer than the 110 ms scan time used in the experiment. Fig.\ref{fig:f4}(a)
shows the simulated values of the concurrence as triangles; their
evolution during the scan is quite similar to that of the experimental data
points.

To assess the quality of the adiabatic scan, we also determined the
fidelities $F_{\exp }=F(\rho _{\exp})$ from the tomographically
reconstructed density operators. The fidelities, which are shown at the top
of Fig.\ref{fig:f4}(a), deviate from unity when the system passes through
the critical points and shows some overall decrease due to decoherence.
Again, the simulated fidelities agree remarkably well with the experimental
values.

As a second order parameter, we also determined the two-spin correlation 
\cite{Sacbook} $\left\langle \sigma _{z}^{1}\sigma _{z}^{2}\right\rangle
=Tr(\rho _{\exp}\sigma _{z}^{1}\sigma _{z}^{2})$, which are shown in Fig.\ref
{fig:f4}(b). As expected, the system is ferromagnetically ordered ($%
\left\langle \sigma _{z}^{1}\sigma _{z}^{2}\right\rangle =+1$) at high
fields, but turns to an antiferromagnetic state ($\left\langle \sigma
_{z}^{1}\sigma _{z}^{2}\right\rangle =-1$) at low fields between the two
quantum critical points. Comparing Fig. \ref{fig:f4}(a) with (b), the
concurrence has the similar behavior to the two-spin correlation.

In conclusion, we have discussed an experimental quantum simulation of a
quantum phase transition in a Heisenberg spin chain. Heisenberg spin chains,
which have been investigated in detail in solid state physics, play an
important role in a number of proposed solid state quantum computers. During
the course of the simulation, the system ground state changes from a
classical product state to an entangled state and back to another product
state. Like in many other proposed quantum simulations, this system had to
be swept adiabatically through the relevant parameter space by properly
varying a Hamiltonian parameter. The techniques developed here may also be
useful for other types of adiabatic quantum computing which have been proved
to be effective for NP-hard problems \cite{Lato2004}, e.g., the classically
NP-hard ground state search. Furthermore, the adiabatic passage can provide
a novel method to create entanglement \cite{Una01}. The simulation can be
extended to other types of Heisenberg spin chains, e.g., Heisenberg XY or
XYZ models etc.; work in this direction is under way.

We thank Dr. X. Zou and Dr. C. Lee for helpful discussions. The experiments
were performed at the Interdisciplinary Center for Magnetic Resonance. X. P.
is supported by the Alexander von Humboldt Foundation. J. D. acknowledges
the support from NUS Research Project (Grant No.R-144-000-071-305) and
National Fundamental Research Program (Grant No. 2001CB309300).

\end{document}